\documentclass[%
 aip,
  amsmath,amssymb,
  preprint,%
]{revtex4-1}

\usepackage{graphicx}
\usepackage{dcolumn}
\usepackage{bm}
\usepackage[utf8]{inputenc}
\usepackage[T1]{fontenc}
\usepackage{mathptmx}
\usepackage{etoolbox}
\usepackage{verbatim}
\usepackage{booktabs}

\makeatletter
\def\@email#1#2{%
 \endgroup
 \patchcmd{\titleblock@produce}
  {\frontmatter@RRAPformat}
  {\frontmatter@RRAPformat{\produce@RRAP{*#1\href{mailto:#2}{#2}}}\frontmatter@RRAPformat}
  {}{}
}%
\makeatother
\newcommand{\bk}{\mathbf{k}}
\newcommand{\bx}{\mathbf{x}}

\begin{document}

\title{Wave-by-wave forecasts in directional seas using nonlinear dispersion corrections}

\author{Eytan Meisner}
\affiliation{The Nancy \& Stephen Grand Technion Energy Program (GTEP), Technion -- Israel Institute of Technology, 3200003 Haifa, Israel}%
\author{Mariano Galvagno}
\affiliation{Centre for Mathematical Sciences, University of Plymouth, PL4 8AA Plymouth, UK}%
\author{David Andrade}
\affiliation{Centre for Mathematical Sciences, University of Plymouth, PL4 8AA Plymouth, UK}
\affiliation{School of Engineering, Science and Technology, Universidad del Rosario, 111711 Bogot\'{a}, Colombia}%
\author{Dan Liberzon}
\affiliation{Faculty of Civil \& Environmental Engineering, Technion -- Israel Institute of Technology, 3200003 Haifa, Israel}%
\author{Raphael Stuhlmeier$^*$}
\affiliation{Centre for Mathematical Sciences, University of Plymouth, PL4 8AA Plymouth, UK}%
\email{raphael.stuhlmeier@plymouth.ac.uk}

\begin{abstract}
    We develop a new methodology for the deterministic forecasting of directional ocean surface waves, based on nonlinear frequency corrections. These frequency corrections can be pre-computed based on measured energy density spectra, and therefore come at no additional computational cost compared to linear theory. The nonlinear forecasting methodology is tested on highly-nonlinear, synthetically generated seas with a variety of values of average steepness and directional spreading, and shown to consistently outperform a linear forecast.  
\end{abstract}

\maketitle

\section{Introduction}
\label{sec:Introduction}

With recent developments in remote sensing of the sea-surface\cite{Ardhuin2019}, coupled with increases in computing power and the growing global role of the maritime economy, deterministic wave forecasting has seen a surge of interest in recent years. Deterministic forecasting models have historically lagged behind the well-developed stochastic forecasting models, which are in daily operational use by national agencies and a large community of stakeholders. However, over the last decade interest has grown in these close-to-real-time predictions which provide wave-by-wave information.

Such predictions are key to maximising power capture from wave energy converters (WECs). While the size and geometry of a particular WEC design is fixed and based on long-term characteristics of the deployment site \cite{Liu2022,Stuhlmeier2018}, it is possible to adapt the damping characteristics to control device motions and thus increase energy capture \cite{Monk2015a,Jin2022}. A key to an effective control strategy is knowing in advance which waves the device will encounter \cite{Zhang2022}, and so relies on wave-by-wave forecasting.

Likewise, offshore operations -- from offloading to the servicing of marine infrastructure -- must often be undertaken close to operational limits. In such cases prior knowledge of wave-induced ship and device motions are beneficial for ensuring safety and prevention of environmental disasters, again calling for accurate monitoring and wave-by-wave forecasting \cite{Ma2018}. Implemented in automated navigation of maritime vessels, such forecasting can also facilitate fuel efficiency and hence reduction of operational costs and associated environmental impacts.  Numerous methodologies for wave forecasting applied to ship motions have been reported in recent years \cite{Connell2015,Kusters2019,Al-Ani2020,Belmont2014}.

The first step in constructing a deterministic (phase-resolved) forecast is to obtain a snapshot of the sea state at a given time. In practice this is usually obtained by remote sensing \cite{reichert1999wamos, belmont2007shallow}. For a given snapshot of the free surface, Fourier transform techniques are commonly used to prepare the forecast. In the simplest implementation of such a forecast, each Fourier mode $k_i$ corresponds to a linear wave mode. Inserting the linear dispersion relation, it is thus possible to ``propagate'' the sea forward, mode-by-mode, in space and time. Such linear forecasts, first investigated in the 1970s \cite{sand1979three}, are robust, and can be expected to perform well in calmer seas \cite{hilmer2014deterministic,Al-Ani2020}. However for  sea-states characterized by higher steepness the initially weak effects of nonlinearity grow in importance. These include energy exchange between modes, an amplitude-dependent dispersion (affecting the propagation speed of the modes) and changes to the wave shape (the appearance of bound modes), which all significantly affect predictions.

It is possible to improve forecast fidelity using numerical methods. For example, the high order spectral (HOS) method \cite{Klein2019} or specific model equations like the nonlinear Schr\"odinger equation \cite{Klein2020} can be used to propagate the measured wave field forward in space and time, at the expense of performing numerical computations. Because typical forecasts have a temporal range of several minutes at best, it is imperative that these can be produced in a timely manner. Moreover, the wave forecasting (or prediction) step is, in practice, only one constituent of a purpose-built system. Forecasting speed is therefore essential.

Here we show how incorporating an algebraic nonlinear frequency correction -- analogous to Stokes' correction for a plane wave -- dramatically improves deterministic forecasts when compared with linear theory alone. This means employing the wave energy spectrum to calculate the dispersion relation of all wave components algebraically, using pre-computed interaction coefficients. This novel approach carries no computational cost compared to linear theory during the forecasting step, and yet compares favourably with directional, nonlinear HOS simulations of synthetic sea-states for moderately long times, depending on the extent of nonlinearity of the sea-state.  In Section \ref{sec:Theoretical background} we discuss the theoretical background of the new method, and provide details of its implementation in Section \ref{sec:Implementation}. We compare linear and nonlinear forecast results obtained using a variety of synthetically generated, highly nonlinear seas in Section \ref{sec:Comparison with HOS}. A discussion of our results and conclusions are found in Section \ref{sec:Discussion and Conclusions}. Supplementary data tables are available in Appendix \ref{sec:tables}.

\section{Theoretical background}
\label{sec:Theoretical background}

\subsection{The 2D discrete Fourier transform}
\label{ssec:The 2D discrete Fourier transform}

The new deterministic forecasting framework is based on Fourier transforms of the sea surface. To this end we review some fundamentals of the 2D discrete Fourier transform, and introduce some notation.
For an $m \times n$ matrix $X,$ which we think of as our input data, i.e.\ a rectangular region of the ocean surface, the discrete Fourier transform is an $m \times n$ matrix $Y$

\begin{equation}
\label{eq: 2D DFTa}
Y_{p+1,q+1} = \sum_{j=0}^{m-1} \sum_{k=0}^{n-1} e^{-2\pi i jp / m} e^{-2 \pi i kq /n} X_{j+1,k+1}
\end{equation}
with inverse transform

\begin{equation}
\label{eq: 2D IDFTa}
X_{p+1,q+1} = \frac{1}{mn}\sum_{j=0}^{m-1} \sum_{k=0}^{n-1} e^{2\pi i jp / m} e^{2 \pi i kq /n} Y_{j+1,k+1}
\end{equation}
where $0 \leq p \leq m-1$ and $0 \leq q \leq n-1.$ If $X$ comes from a discretisation of an area  $L_x \times L_y$ of the sea surface into $m \times n$ samples we can write compactly 
\[ x_j = jL_x/m \text{ and } y_k = kL_y/n, \]
and so $2 \pi i j p /m = 2 \pi i p x_j / L_x$ and $2 \pi i k q /n = 2 \pi i q y_k / L_y.$ The definitions
\[ \mu_p = 2 \pi p / L_x, \, \nu_q = 2 \pi q / L_y \]
allow for the reformulations 
\begin{equation}
\label{eq: 2D DFTb}
Y_{p+1,q+1} = \sum_{j=0}^{m-1} \sum_{k=0}^{n-1} e^{-i x_j \mu_p} e^{-i y_k \nu_q} X_{j+1,k+1}
\end{equation}
and
\begin{equation}
\label{eq: 2D IDFTb}
X_{p+1,q+1} = \frac{1}{mn}\sum_{j=0}^{m-1} \sum_{k=0}^{n-1} e^{i x_p \mu_j} e^{i y_q \nu_k} Y_{j+1,k+1}.
\end{equation}

Using trigonometric interpolation polynomials we can introduce continuous variables 
\begin{equation}
\label{eq: 2D IDFT cont}
X(x,y) = \frac{1}{mn}\sum_{j=-m/2}^{m/2-1} \sum_{k=-n/2}^{n/2-1} e^{i (x \mu_j+ y \nu_k)} Y_{j+1,k+1}.
\end{equation}
where we recognize the wavevector $\bk = (\mu_j, \nu_k)$ and position $\bx = \left(x,y\right)$ in the term $\exp(i\bx\cdot\bk).$ 

\subsection{Linear forecasts and the predictable region}
\label{ssec:Linear forecasts and the predictable region}

For a wavenumber vector $\bk=(\mu,\nu)$ with $k=\|\bk\|$  the linear dispersion relation in deep water reads 
\begin{equation}
\label{eq: lin disp rel}
\omega^2 = gk.
\end{equation} 
The energy of a wave with wavenumber $\bk$ propagates at the group velocity $c_g.$ 
Writing $\bk = (k \cos(\theta), k \sin(\theta))$ we can divide the group velocity into components $(c_{g,x},c_{g,y}) = (c_g \cos(\theta), c_g \sin(\theta)) = (d\omega/d\mu, d\omega/d\nu).$

This information is sufficient to determine how each Fourier mode in the sea surface \eqref{eq: 2D IDFT cont} propagates, and so to produce a linear wave forecast with time $t$. Writing  $\omega(\bk)$ for the frequency of the corresponding mode, we obtain from the Fourier transform a surface that evolves in space and time:
\begin{equation}
\label{eq: linear forecast}
\Tilde{\eta}_L(\bx,t) = \frac{1}{mn}\sum_{j=-m/2}^{m/2-1}\sum_{k=-n/2}^{n/2-1} e^{i (x \mu_j+ y \nu_k - \omega(\mu_j,\nu_k)t)} Y_{j+1,k+1}.
\end{equation}

\begin{figure*}[h!]
\centering
\includegraphics[width=0.9\linewidth]{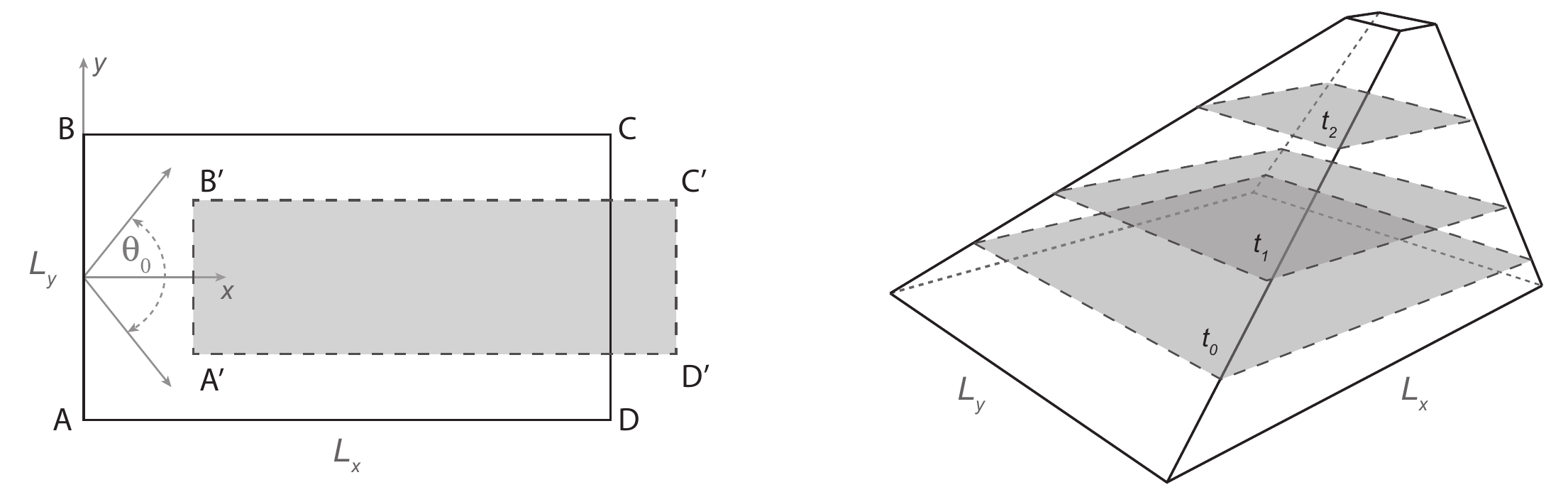}
\caption{Schematic of the evolution of the predictable region in $(x,y)$-space (left panel) from time $t=0$ (solid rectangle) to time $t=t_0>0$ (dashed rectangle). The principal direction of wave propagation is the $x-$direction, and directional spread is limited to $\theta\in[-\theta_0,\theta_0],$ for $\theta_0 < \pi/2.$ The evolution in $(x,y,t)$ is shown indicatively in the right panel.}
\label{fig: Predictable region 2D}
\end{figure*}

\begin{figure}[h!]
\centering
\includegraphics[width=0.9\linewidth]
{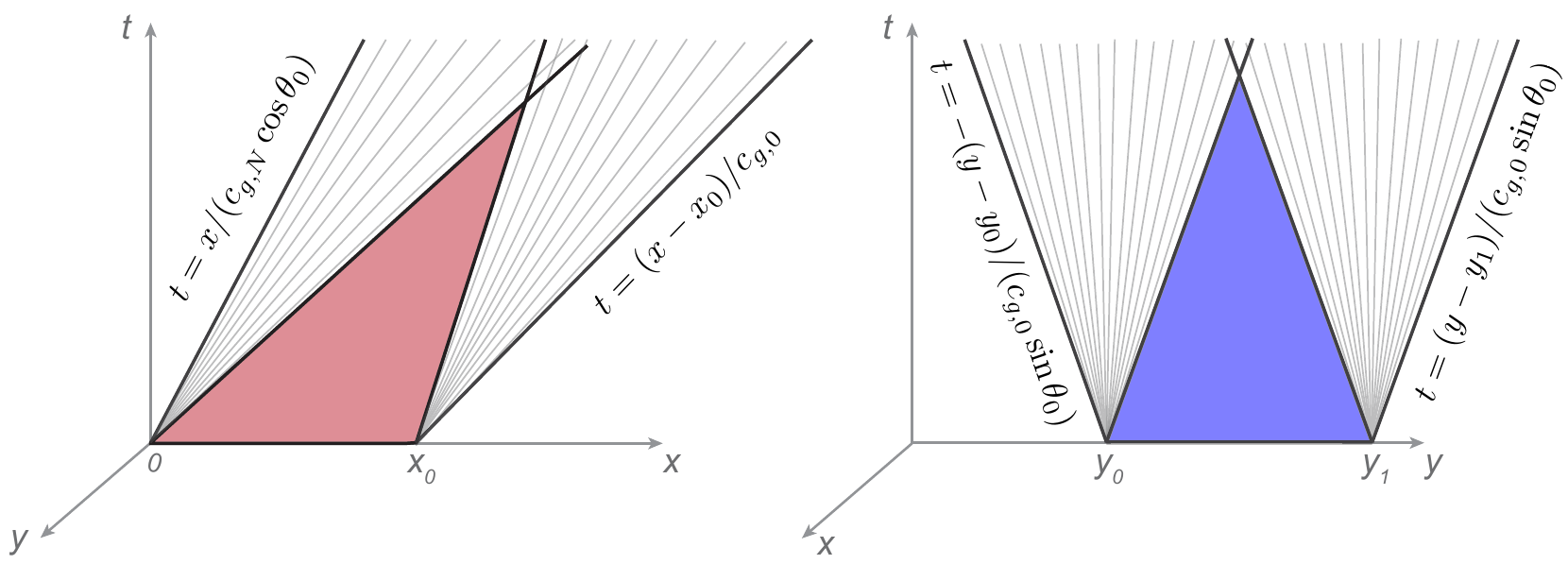}
\caption{Sections through the 3D predictable region highlighting the fastest and slowest wave rays advancing in the $x$-direction. (Left panel) Predictable wedge in the $(x,t)$-domain, bounded by rays $t \sim x/(c_{g,N}\cos\theta_0)$ and $t\sim x/c_{g,0}.$ (Right panel) Predictable wedge in the $(y,t)$-domain, bounded by rays $t \sim \pm y/(c_{g,0} \sin\theta_0).$}
\label{fig:PR-Section}
\end{figure}

The dispersive nature of the waves dictates the region over which prediction is possible -- its size depends on the initially measured area, and the lengths and directions of the waves present \cite{Naaijen,Wu2004a}. Assuming the magnitudes of the wavenumber vectors $k\in [k_0,k_N]$ (where $k_0$ and $k_N$ are the longest and shortest waves resolved in the record, respectively; see Section \ref{sec:Implementation} below on the role of these cut-offs) and the propagation angles $\theta \in [-\theta_0,\theta_0]$ for $\theta_0 < \pi/2,$ we can show how the predictable region evolves with time (see Figures \ref{fig: Predictable region 2D} \& \ref{fig:PR-Section}).

Fundamentally, the energy of a mode $\bk = (\mu,\nu)$ at $(x_0,y_0)$ travels in time $t$ to a location $(x_1,y_1)$ with
\begin{align*}
&x_1 = x_0 + c_g t \cos(\theta),\\
&y_1 = y_0 + c_g t \sin(\theta).
\end{align*}
Here $\theta = \arctan(\nu/\mu).$ Therefore, if the initially measured area is rectangular with Cartesian coordinates $A=(x_a,y_a), \, B=(x_b,y_b), \, C=(x_c,y_c)$ and $D=(x_d,y_d),$ at a later time $t=t_0$ the vertices of the predictable region will be 
\begin{subequations}
\label{eq:predictable region at t0}
\begin{align}
&A' = (x_a + c_{g,0} t_0, y_a + c_{g,0} \sin(\theta_0)t_0),\\
&B' = (x_b + c_{g,0} t_0, y_b - c_{g,0} \sin(\theta_0)t_0),\\
&C' = (x_c + c_{g,N} \cos(\theta_0) t_0, y_c - c_{g,0} \sin(\theta_0)t_0),\\
&D' = (x_d + c_{g,N} \cos(\theta_0) t_0, y_d + c_{g,0} \sin(\theta_0)t_0).
\end{align}
\end{subequations}

The principal direction of wave propagation is along the $x$-axis, and the line $\overline{A'B'}$ in Figure \ref{fig: Predictable region 2D} shows how far the fastest modes have travelled from $\overline{AB}$ in time $t_0$, while $\overline{C'D'}$ shows how far the slowest  have travelled from $\overline{CD}$ in the same time. 

By the same token, we need to capture obliquely propagating waves in the positive and negative $y$-directions, as shown in Figure \ref{fig:PR-Section}. The value of $\sin(\theta)$ is largest/smallest for $\theta=\theta_0/-\theta_0,$ since we are restricted to $\theta_0 < \pi/2.$ This means the predictable region is bounded by rays $t \sim \pm y/(c_{g,0} \sin(\theta_0)),$ as shown in Figure \ref{fig:PR-Section}.

The maximum temporal extent of the prediction can then be given by
\begin{equation} \label{eq:t-predictable} t_\infty = \min \left\lbrace t^{(x)}_\infty, t^{(y)}_\infty \right\rbrace. \end{equation}
where 
\[ t^{(x)}_\infty = \frac{L_x}{c_{g,0} - c_{g,N} \cos(\theta_0)}, \quad t^{(y)}_\infty = \frac{L_y}{2 c_{g,0} \sin(\theta_0)}. \]
Here $L_x$ and $L_y$ are the measured distances in the $x$ and $y$ directions, respectively.
A visualisation of such a predictable region in terms of $(x,y,t)$ is given in Figure \ref{fig: Predictable region 2D} (right panel). 

\subsection{Nonlinear forecasts}
\label{ssec:Nonlinear forecasts}

The linear theory presented above is easy to understand and implement. While waves which are not too steep can be modelled with sufficient accuracy using the linear theory, it has been recognised since early work by G.~G.\ Stokes that physical water waves have a dispersion relation which depends on their amplitude -- in general, steeper waves travel faster. Moreover, not only does the amplitude of a wave itself influence its dispersion relation, but waves in the sea have a mutual influence on one another as elucidated in work of Longuet-Higgins and Phillips \cite{Longuet-Higgins1962} for two wave trains. This mutual, nonlinear dispersion correction to third order in wave steepness, has subsequently been derived in the framework of the Zakharov equation \cite{Stuhlmeier2019}, where the corrected frequency is written
\begin{equation} \label{eq: discrete nonlinear dispersion}
{\Omega_s = \omega_s + \sum_r e_{sr} T_{srsr} |B_r|^2.}
\end{equation}
Here $e_{sr} = 1$ for $s = r$ and $e_{sr} = 2$ for $s \neq r,$ and $|B_r|$ denotes the modulus of the complex amplitude $B(\bk_r,t=0)$ of the Zakharov equation (see \onlinecite[Eq.\ (14.2.13)]{Mei2005}), which is related to the Fourier transform of the free-surface elevation and the potential at the free surface. The crucial ingredients in this compact formulation are the interaction kernels $T_{ijkl}=T(\bk_i,\bk_j,\bk_k,\bk_l)$.  The full expressions for these are algebraically cumbersome, but in deep water their symmetric form (when indices $i=k$ and $j=l$, or $i=l$ and $j=k$) can be simplified to  \cite{Leblanc2009}
\begin{align*} 
 T(\bk_i, \bk_j, \bk_i, \bk_j)  &= - \frac{1}{16 \pi^2 (|\bk_i| |\bk_j|)^{1/2}} \Biggl[ 3(|\bk_i| |\bk_j|)^2 \\ 
& \left. + (\bk_i \cdot \bk_j)(\bk_i \cdot \bk_j - 4 (|\bk_i| + |\bk_j|)(|\bk_i| |\bk_j|)^{1/2}) \right. \\
& + \frac{2(\omega_i - \omega_j)^2 (\bk_i \cdot \bk_j + |\bk_i| |\bk_j|)^2}{g |\bk_i - \bk_j| - (\omega_i - \omega_j)^2} \\
& + \frac{2(\omega_i + \omega_j)^2 (\bk_i \cdot \bk_j - |\bk_i| |\bk_j|)^2}{g |\bk_i + \bk_j| - (\omega_i + \omega_j)^2} \Biggr].
\end{align*}
The complex amplitudes $|B_r|$ are directly related to the measured Fourier amplitudes via 
\begin{equation} \label{eq:B_p-Y-relation} 
|B_r| = \frac{2 \pi}{mn} \sqrt{\frac{2g}{\omega_r}} |Y_r|,
\end{equation}
and enter into the nonlinear corrected frequencies \eqref{eq: discrete nonlinear dispersion}. This leads naturally to a change in the group velocities, and generally to a small growth in the predictable region \cite{Galvagno2021}.  
We note that these frequency corrections are exact to third order, in contrast to the average nonlinear dispersion correction implemented by Desmars et al \cite{Desmars2020}. They recover the classical results for one and two modes \cite{Stokes1847,Longuet-Higgins1962}, as detailed in Stuhlmeier \& Stiassnie\cite{Stuhlmeier2019}.

\section{Implementation of linear and nonlinear forecasts}
\label{sec:Implementation}

The starting point for an implementation of the forecasting methodology described above is a measurement of the sea-surface over an area $L_x \times L_y.$ The measurement resolution  $m \times n$ should be such that the waves of interest (for example, those waves close to the spectral peak) can be resolved accurately. In practical applications the resolution may be dictated by the measurement technology used, and it may prove necessary to assimilate data from multiple measurements or devices (e.g.\ if arrays of buoys are used\cite{Hlophe2023}) in order to reconstruct the free-surface elevation with the necessary or desired resolution. Since our methodology aims to improve the prediction step, we will assume that a suitable record is available. Spectral information about the discretised record can then be efficiently extracted using the Fast Fourier Transform (FFT).

The Fourier transformed signal is used to determine the predictable domain. It is also necessary to choose cut-offs $k_0$ and $k_N$ (note that these are distinct from the smallest/largest wavenumbers which can be resolved based on the discretisation) for the longest and shortest waves of interest --  other modes being discarded from the forecast. If $k_0$ is chosen too small and $k_N$ too large, the effective prediction time \eqref{eq:t-predictable} may be too short for the desired application. Very small $k_0$, for example, will lead to a  large group velocity $c_{g,0},$ and a decrease in the slopes of the rays along which mode-$k_0$ energy propagates in Figure \ref{fig: Predictable region 2D}. This yields a shrinking of the predictable regions (indicated in red and blue in the top and bottom panel, respectively), and a smaller value of $t_{\infty}.$  On the other hand, if the cut-offs are too generous then appreciable energy will be lost, extending the prediction time at the expense of forecast accuracy. 

Various practical possibilities exist in the choice of such cut-offs (see e.g.\ the discussion in Desmars et al \cite[Sec.\ 3.3]{Desmars2020}). While it is possible to use, for example, fixed multiples of the peak wavenumber $k_p$, an adaptive method that takes the spectral shape into account is preferable. One method, which we employ below, is to use a cut-off which eliminates modes whose energy is less than a fixed percentage of the spectral peak energy. 

\begin{figure}[h!]
    \centering
   \includegraphics[scale=0.30]{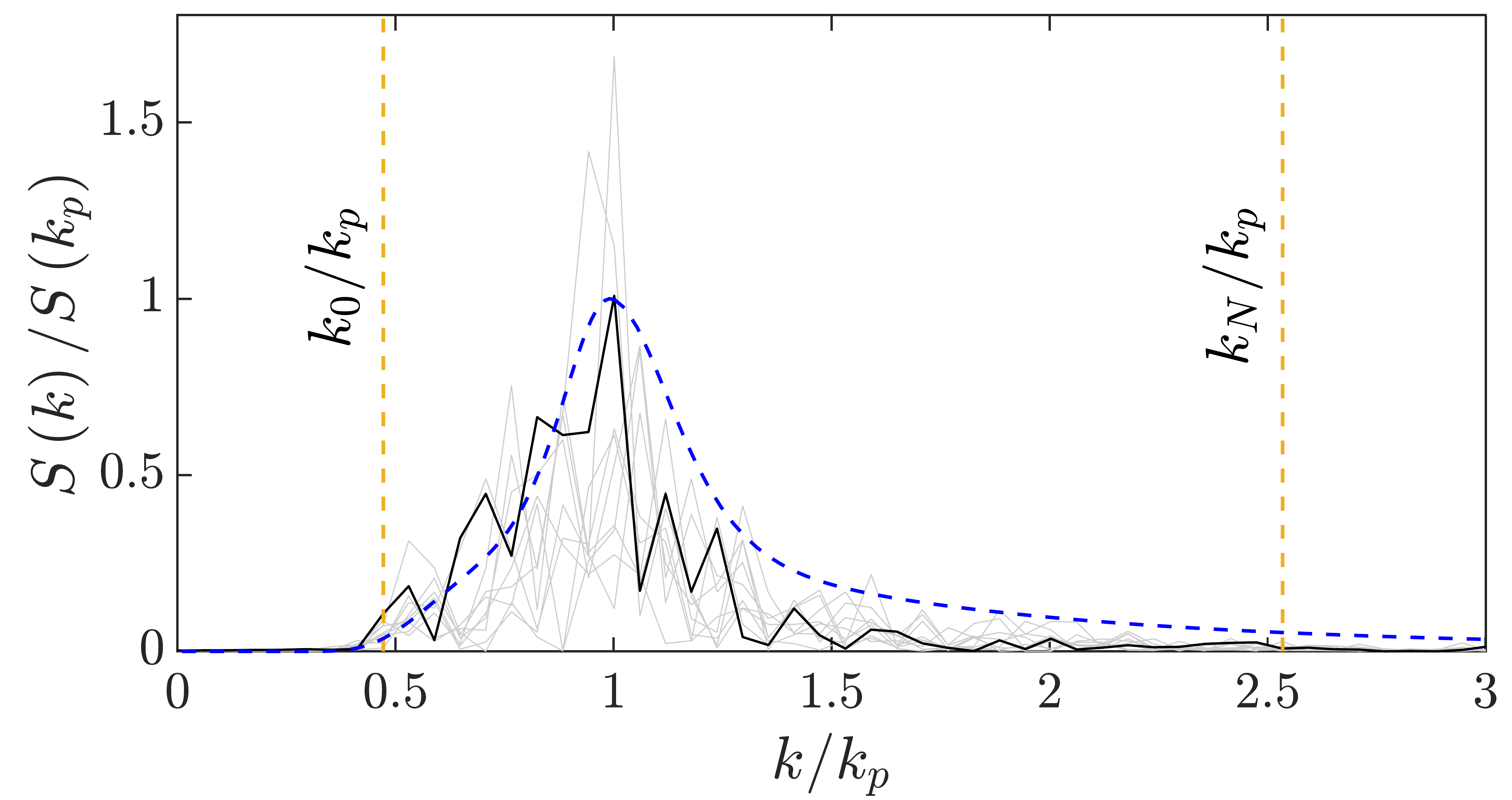}
    \caption{Averaged variance density spectrum (blue, dashed curve) and raw variance density spectrum (black, solid curve) computed from a single amplitude per frequency (further realisations of raw variance density spectra are plotted in light grey). Yellow dashed vertical lines show cut-off values $k_0$ and $k_N$ which represent 1\% of the spectral peak energy of the averaged variance density spectrum.}
    \label{fig:Spectral-cut-offs-instantaneous-vs-averaged}
\end{figure}

Finding robust wavenumber cut-offs is made more difficult when the spectrum  employed comes from a single measurement, and thus has the ``grassy” shape typically seen when there is only a single amplitude per wavenumber (i.e.\ the grey and black curves in Figure \ref{fig:Spectral-cut-offs-instantaneous-vs-averaged}). It is thus advantageous to compute a variance density spectrum (or energy density spectrum) from multiple realisations (or, via windowing, from a single realisation, see Holthuijsen\cite[App.\ C]{Holthuijsen2007}), shown as the blue, dashed line in Figure \ref{fig:Spectral-cut-offs-instantaneous-vs-averaged}. This smoothed spectrum makes the identification of suitable cut-off values simpler and stabler. While the energy spectrum also changes, such changes occur on a much slower time-scale, which may range from 30 minutes during storms to several hours during more quiescent periods\cite{Fedele2013,Fedele2019a,Knobler2022}. In contrast, the practical range of each deterministic forecast under typical oceanic conditions is only 1-2 minutes. The established wavenumber cut-offs can therefore be re-used for multiple predictions.

In order to implement the nonlinear frequency corrections \eqref{eq: discrete nonlinear dispersion}, two further computations are necessary: the most costly of these is the evaluation of the $m^2 \times n^2$  interaction kernels $T_{srsr}$ for the $m \times n$ modes. However, since these kernels depend only on the chosen discretisation, but not on any features of the sea-surface, they can be easily pre-computed and stored. 

The second ingredient in the nonlinear forecast consists of the complex amplitudes $|B_r|^2$. These are simply re-scaled Fourier amplitudes, and could be obtained directly from the measurement at time $t=0,$ i.e.\ the instantaneous amplitude spectrum, with a shape akin to the grey/black curves in Figure \ref{fig:Spectral-cut-offs-instantaneous-vs-averaged}. A more stable value for these complex amplitudes can be obtained by extracting them from the averaged variance density spectrum, i.e.\ the blue, dashed curve in Figure \ref{fig:Spectral-cut-offs-instantaneous-vs-averaged}. We denote this
\begin{equation} S(\mu,\nu) \approx \frac{1}{\Delta \mu \Delta \nu} E \left\lbrace \frac{1}{2}\mathbf{a}^2 \right\rbrace, \end{equation}
where $E$ on the right denotes the expected value, $\Delta \mu, \, \Delta \nu$ denote the discretisation in wavenumber space, and $\mathbf{a}$ is the vector of Fourier amplitudes. The relation between \eqref{eq:B_p-Y-relation} and the amplitude $a_r$ corresponding to the $r$-wave vector is given by
\[ | B_r |^2 = \frac{4 g \pi^2}{\omega_r} \frac{a_r^2}{2}.\]
Hence, the average value of $|B_r|^2$ is
\[ E\{|B_r|^2\}=\frac{4 g \pi^2}{\omega_r} E\left\lbrace\frac{a_r^2}{2}\right\rbrace=\frac{4 g \pi^2}{\omega_r} S(\mu,\nu)\Delta\mu\Delta\nu,\] 
where the wave vector $\bk = (\mu,\nu)$ and $S$ is the energy spectrum as a function of wave vector. 

Using the averaged variance density spectrum (or energy spectrum) eliminates the many fluctuations inherent in the instantaneous amplitude spectrum, see Figure \ref{fig:Spectral-cut-offs-instantaneous-vs-averaged}.  Once sufficient data has been gathered to generated a well-resolved energy spectrum, this provides a stable input for calculating the corrected dispersion relation \eqref{eq: discrete nonlinear dispersion}, and producing forecasts. This approach also avoids unnecessary recalculation of the nonlinear corrected frequencies.

\section{Comparison with HOS simulations}
\label{sec:Comparison with HOS}
\subsection{Generation of synthetic seas via HOS}
\label{ssec:Generation of synthetic seas via HOS}

In order to test the above forecasting methodology we will generate synthetic, directional wave fields by means of the higher-order spectral method, using the open source HOS--Ocean code \cite{Ducrozet2016}. These wave-fields are initialised by a directional JONSWAP spectrum (here written in terms of frequency $\omega$ and direction $\theta$)
\begin{equation} 
S\left(\omega,\theta\right)=F\left(\omega\right)\cdot G\left(\theta\right)
\end{equation}
where $F\left(\omega\right)$ is the JONSWAP spectrum specified by the peak period $T_{p}=2\pi/\omega_p$, the significant wave height  $H_{s}$ and shape factor $\gamma$. 
\begin{equation}
F\left(\omega\right)=\alpha_J H^2_s\omega^4_p\omega^{-5}\exp\left[-\frac{5}{4}\left(\frac{\omega}{\omega_p}\right)^{-4}\right]\gamma^{\exp\left[-\frac{\left(\omega-\omega_p\right)^2}{2\sigma^2\omega^2_p}\right]},
\end{equation}
with $\sigma \in \left( 0.07\,\,\text{for}\,\,\omega<\omega_p; 0.09\,\,\text{for}\,\,\omega\ge\omega_p\right)$ and $\alpha_J$ chosen to obtain the desired significant wave height.
$G\left(\theta\right)$ in turn is the directional  spreading function defined as
\begin{equation} \label{eq:D-theta}
G\left(\theta\right) = \begin{cases} \frac{1}{\beta} \cos^2\left({\frac{2\pi\theta}{4\beta}}\right) \text{ for } |\theta|<\beta ,\\ 0 \text{ otherwise.} \end{cases}
\end{equation}
where $\theta$ is the angular (polar) distribution with $\theta\in\left(-\pi,\pi\right)$ and $\beta$ the directionality (see Socquet-Juglard et al \cite[Eq.\ 4 \& Figure 1]{Socquet-Juglard2005}).

\begin{figure}
\centering
\includegraphics[scale=0.7]{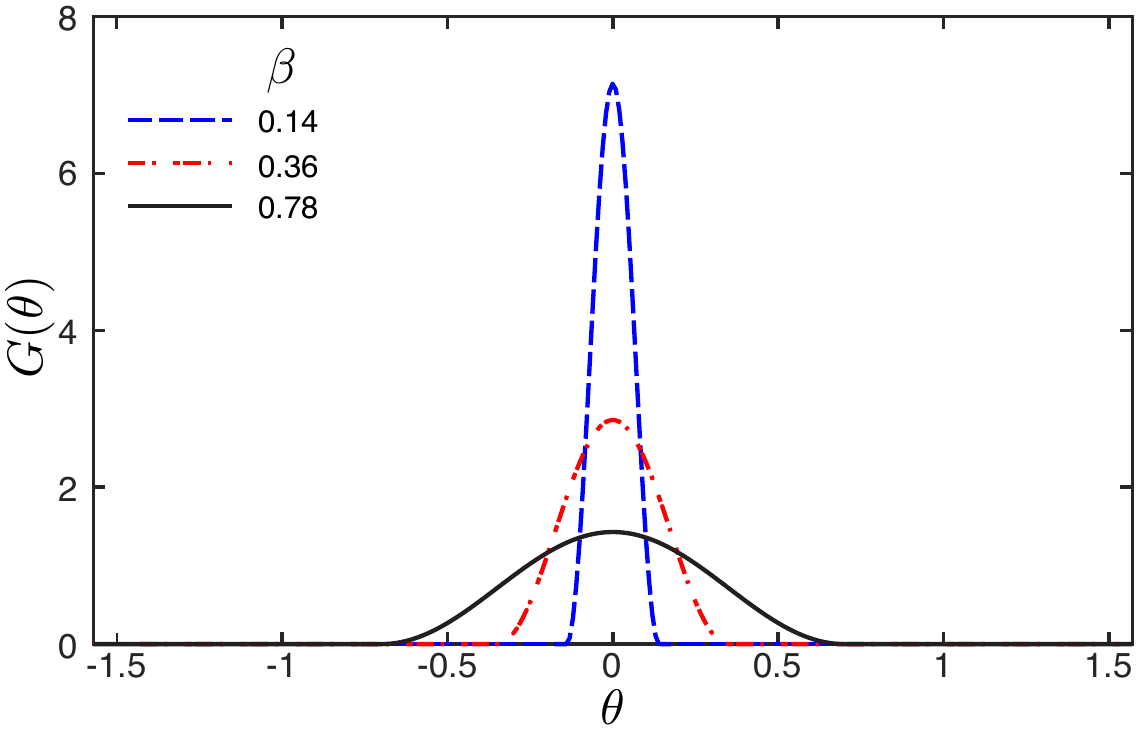}
\caption{Shape of the directional spreading function \eqref{eq:D-theta} for $\beta=0.14, \,0.36$ and 0.78.}
\label{fig:D-theta}
\end{figure}

In all cases considered we generate waves with peak period  $T_{p}=10$ s, i.e.\ $\lambda_p \approx 157.1$ m. To compare linear and nonlinear forecasts, we employ a range of significant wave-heights $H_{s}=$ 3 m, 5 m, 7 m, and 9 m. These correspond to values of characteristic steepness $H_s/\lambda_p$  of 1.9\%, 3.2\%, 4.5\% and 5.7\%, respectively. We further capture the effect of directional spreading by simulating seas with $\beta = 0.14, \, 0.36$ and 0.78 (see Figure \ref{fig:D-theta}). Our JONSWAP shape-factor $\gamma$ is set equal to 3.3 in all cases considered.

\subsection{Comparison of HOS and linear/nonlinear forecasts with identical resolution}
\label{ssec:Comparison Whole Domain}

A significant advantage of synthetically generated data is the ability to precisely differentiate various effects, including the incorporation of correct dispersion characteristics, the presence of nonlinear energy exchange, and the effects of discretisation. In order to eliminate the latter entirely, in this section we describe linear/nonlinear forecasting based on the exact modes used in the HOS simulation. 

The HOS domain of size $L_x = L_y = 2600$ m is initialised with $128 \times 128$ modes, and HOS-Ocean with nonlinearity $M=5$ is used to generate ten realisations of each JONSWAP spectrum as detailed in Section \ref{ssec:Generation of synthetic seas via HOS}, i.e.\ 120 realisations in total, each with a simulation time of 250 s. (Note that the first 100 s are used for the relaxation scheme of the HOS, \cite{Ducrozet2007,Ducrozet2016,Dommermuth2000} with relaxation parameter $n=4.$)

For each realisation the 2D Fourier transform of the domain with $128 \times 128$ Fourier modes is taken to initialise the forecast. This “perfect” resolution means that the initial Fourier spectrum for the linear/nonlinear forecast coincides exactly with that used by the HOS. This also implies that the predictable region does not shrink, since all Fourier modes are accounted for. The mismatch between forecast and the HOS “sea” is therefore attributable entirely to the effects of nonlinearity on wave propagation.

We shall measure quality of fit using two metrics: the linear correlation between sea and forecast $\rho$, and the normalised mean square error (NMSE)
\begin{align}
\label{eq:NMSE}
\mathcal{E}(t_i) =  \frac{{\lVert \eta(x,y,t_i)-\tilde{\eta}(x,y,t_i) \rVert}^2}{{\lVert \eta(x,y,t_i)\rVert}^2}. 
\end{align}
Here $\eta(x,y,t_i)$ is the measured surface elevation, and $\tilde{\eta}(x,y,t_i)$ the predicted surface elevation at time $t_i = 0,\, 30, \, 60, \, 90,$ or 120 s. In practice both $\eta$ and $\tilde{\eta}$ are matrices, and $\lVert\cdot\rVert$ denotes the Frobenius norm, generalising the Euclidean norm. Linear correlation measures simply whether the two signals rise and fall synchronously -- whether we correctly forecast when crests and troughs occur -- and takes on values between -1 and 1 (perfect correlation). NMSE is an aggregate measure of quality that also takes the amplitude into account, and ranges between 0 (perfect fit) to 2 if the signals have the same energy (variance). Both measures vary from realisation to realisation, and the results we show are averages (denoted by a bar) over 10 realisations for both linear correlation $\overline{\rho}$ and NMSE $\overline{\mathcal{E}}$.

\begin{figure}
    \centering
    \includegraphics[scale=0.7]{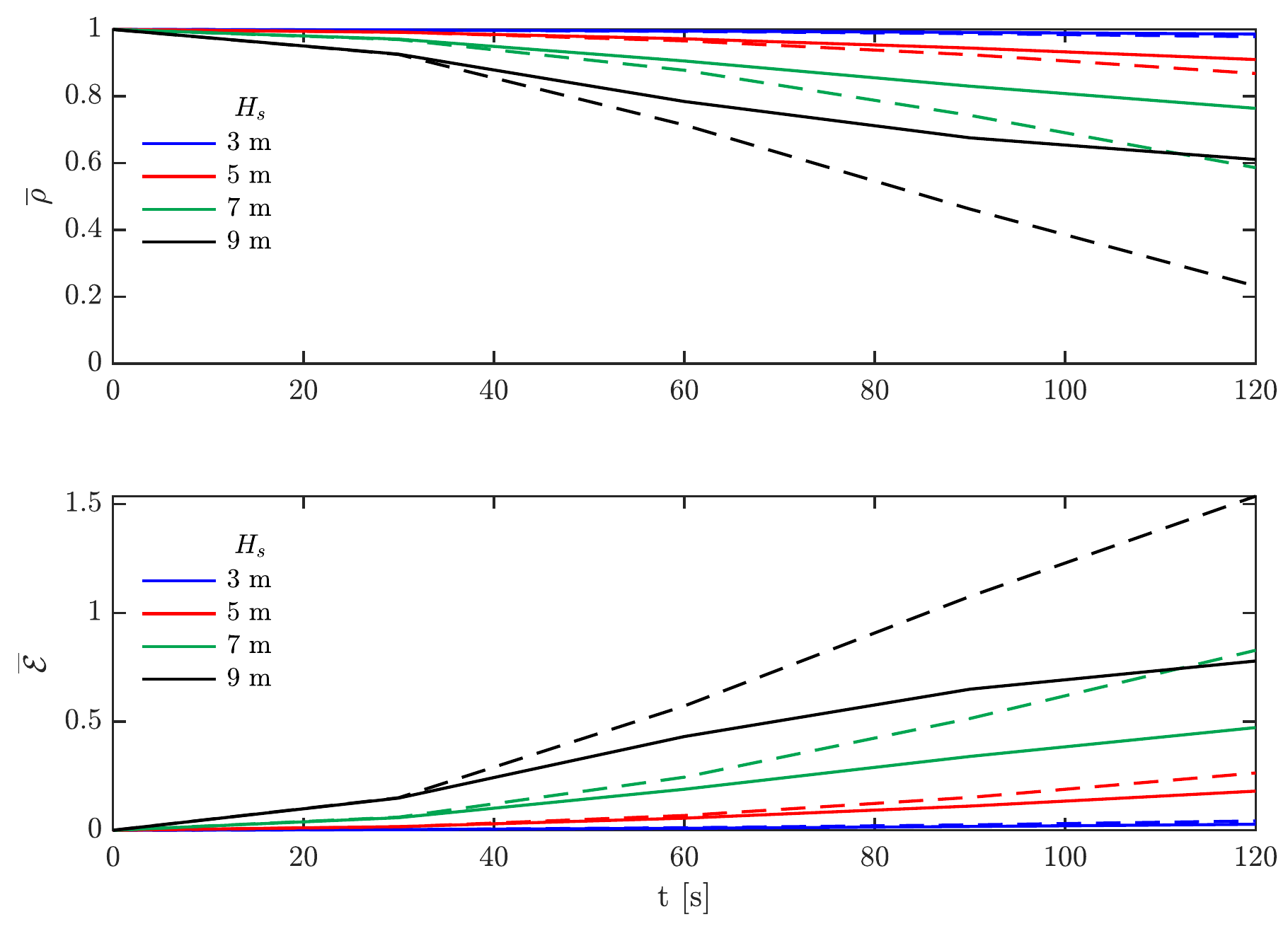}
    \caption{Plots of the averaged linear correlation (top) and NMSE (bottom) as a function of time for significant wave heights $H_s =$ 3, 5, 7 and 9 m. Only $\beta =0.14$ is shown. Dashed lines denote the linear forecast vs. HOS sea. Solid lines denote the nonlinear forecast vs. HOS sea.}
    \label{fig:FullRes-Rho_NMSE}
\end{figure}

For this whole-domain forecast the averaged correlation and NMSE are shown in Figure \ref{fig:FullRes-Rho_NMSE} (results for $\beta = 0.14$ only are plotted; numerical values for $\beta=0.14, \,0.36$ and 0.78 are available in the appendix, Tables \ref{table:Corr Full Domain}--\ref{table:NMSE Full Domain}). In all cases the dashed lines depict the linear forecast, while the solid lines depict the nonlinearly corrected forecast. As expected, for the lowest sea-states with $H_s = 3$ m there is barely a discernible difference between the two forecasts. However, for sea-states of higher steepness values (particularly $H_s=7$ m and 9 m, which correspond to characteristic steepness of $H_s/\lambda_p = 4.5\%$ and 5.7\%, respectively) the differences between linear and nonlinear forecasts beyond 30 s are marked. These results serve as a benchmark for what can be achieved over a shrinking domain with imperfect knowledge, which is treated below.

\subsection{Comparison of HOS and linear/nonlinear forecasts within the predictable region}
\label{ssec:Comparison Predictable Region}

\begin{figure*}
\centering
\includegraphics[width=0.9\linewidth]
{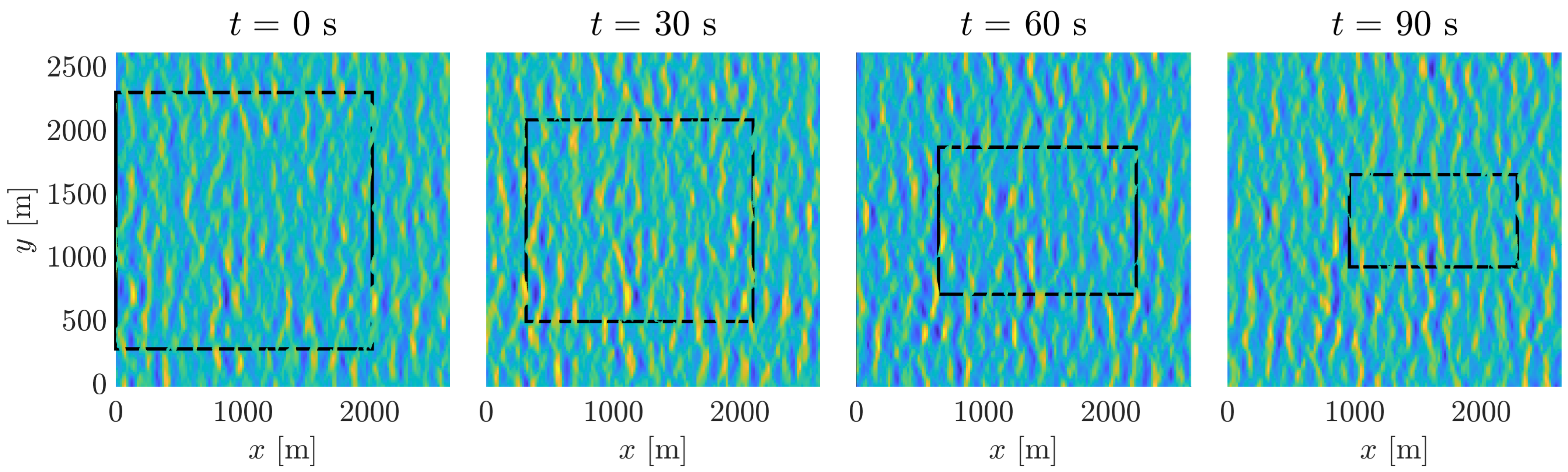}
\caption{Shrinking and advance of the predictable region in time (black dashed line), with respect to an initial region $\eta\left(x,y,t=0\right)$.
}
\label{fig:shrinking_domain}
\end{figure*}

In this section we shall discuss the more realistic case of forecasting within an ocean basin: we select a 2000 m $\times$ 2000 m square of the HOS domain, which should be envisioned as a surface measured at $t=0$ s via remote sensing (see Figure \ref{fig:shrinking_domain}). When it comes to practical implementation we do not know which Fourier modes are present in nature. This circumstance inevitably leads to uncertainties and the neglect of some part of the wave energy, as detailed in section \ref{ssec:Linear forecasts and the predictable region}.  

For the calculation of the predictable region, as well as the amplitudes which enter into the nonlinear frequency corrections, we employ the averaged energy spectrum, as described above. In practice this would mean collecting multiple amplitude spectra from snapshots and averaging these, or using windowing to extract a smooth, averaged energy spectrum (cf.\ the blue, dashed line in Figure \ref{fig:Spectral-cut-offs-instantaneous-vs-averaged}) from a large, single snapshot. Because the underlying spectrum for our synthetic data is known, we shall simply employ the directional JONSWAP spectrum which initialises the HOS. We discretise into a grid of 160 $\times$ 160 points, which is sufficient to resolve waves of interest around the spectral peak. This discretisation differs from the grid size used in the HOS simulations (note that our prediction region is a subdomain of the larger HOS simulation).

While this spectral resolution with $\Delta k\approx 0.003$ is suitable for the waves found in our sea, we must truncate the energy further to establish a useful predictable region. To this end, we search for all modes with less than $1\%$ of the energy density of the spectral peak, and the smallest and largest such modes are chosen as $k_0$ and $k_N,$ respectively (see Section \ref{ssec:Linear forecasts and the predictable region}). These cut-offs in turn specify $c_{g,0}$ and $c_{g,N},$ and the directional spreading $\theta$ is extracted from the $\arctan(\nu/\mu)$  (recall $\mu$ and $\nu$ are the $x-$ and $y-$components of the wavenumber, respectively). The maximum of $\theta$ with non-negligible energy is taken as the maximal directional spreading value $\theta_0.$ Together these provide all the necessary information to determine the predictable region, which is shown in black in Figure \ref{fig:shrinking_domain} at times $t=0$, 30, 60 and 90 s. The chosen cut-offs can be easily implemented, particularly when the averaged energy density spectrum is employed (see Figure \ref{fig:Spectral-cut-offs-instantaneous-vs-averaged}), and lead to a reduction by approximately $5\%$ in the total spectral energy, as detailed in Table \ref{table:spectral-cutoffs}. This is similar to cut-offs implemented by Desmars et al \cite{Desmars2020} or Huchet et al \cite{Huchet2021}.

\begin{table*}
\centering
\begin{tabular}{@{} ccccccc @{}}
$\beta$ [rad] & $\theta_0$ [rad] & $k_0$ [1/m] & $k_N$ [1/m] & $c_{g,0}$ [m/s] & $c_{g,N}$ [m/s] & spectral energy \\
\midrule
0.14 & 0.1419 & 0.0213 & 0.1541 & 10.7206 & 3.9890 & 95.99\%\\
0.36 & 0.3218 & 0.0213 & 0.1543 & 10.7206 & 3.9871 & 95.42\%\\
0.78 & 0.7230 & 0.0212 & 0.1548 & 10.7540 & 3.9807 & 94.95\%\\
\end{tabular}
\caption{Wavenumber cut-offs and associated parameters for the three directional spreading coefficients $\beta=0.14, \, 0.36$ and 0.78. 
Further detail about the calculation of these parameters is given in Section
\ref{ssec:Comparison Predictable Region}.}
\label{table:spectral-cutoffs}
\end{table*}

\begin{figure}
\centering
\includegraphics[width=0.8\linewidth]
{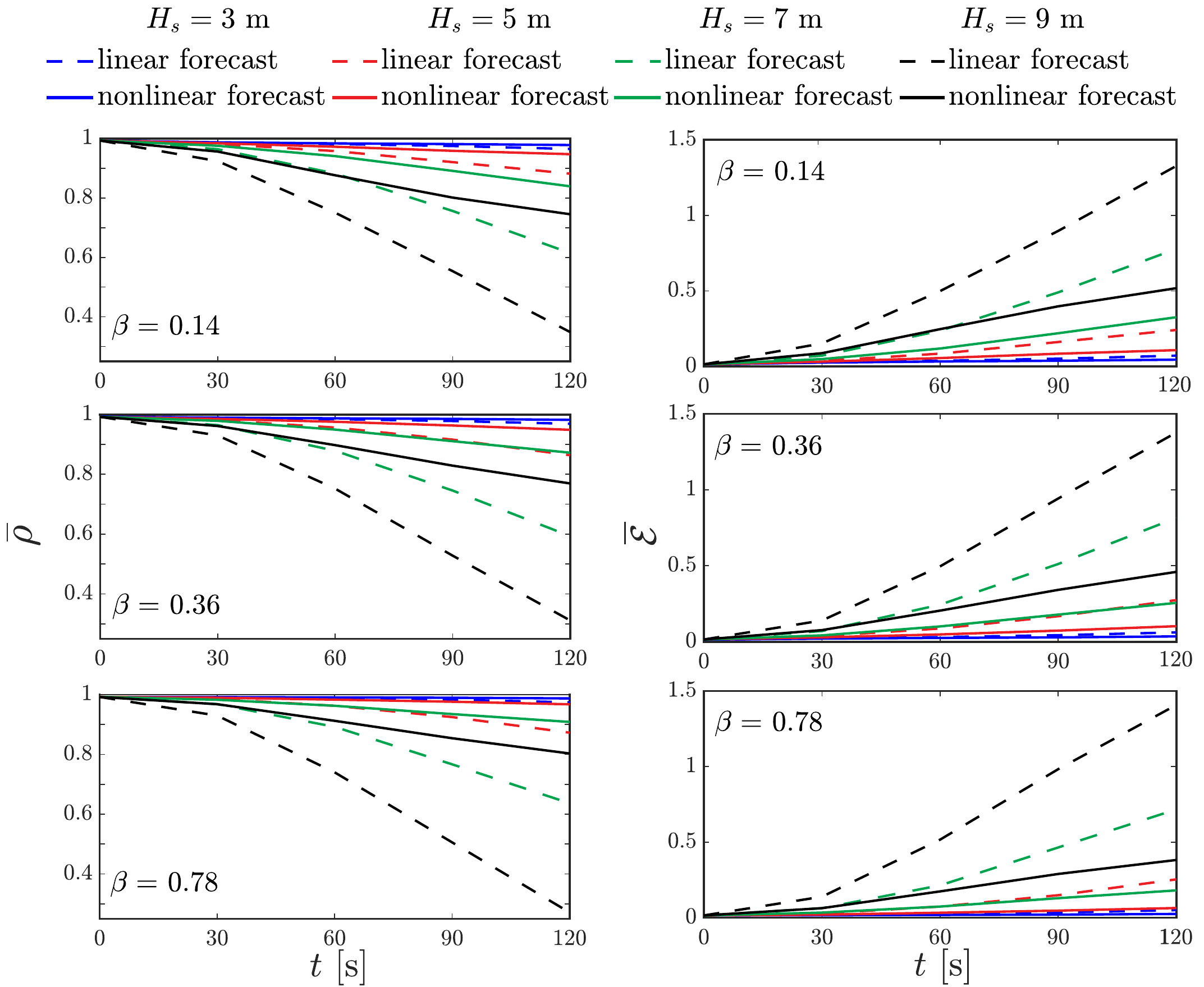}
\caption{Left panel: linear correlation $\overline{\rho}$ coefficients for the linear and nonlinear forecasts as a function of time, each averaged over 10 realisations. Right panel: normalized mean squared error $\overline{\mathcal{E}}$ for the linear and nonlinear forecasts as a function of time, each averaged over 10 realisations.}
\label{fig:corrcoef_beta_nmse_beta}
\end{figure}

A comparison of linear and nonlinear forecasts within the predictable region (calculated on the basis of the linear group velocities) is shown in the {left panel and right panel of Figure \ref{fig:corrcoef_beta_nmse_beta}, respectively}. As above, dashed lines denote the linear forecasts, while solid lines denote the nonlinear corrected forecasts. 
{The left panels of Figure \ref{fig:corrcoef_beta_nmse_beta}} show the averaged linear correlation for all three values of directional spreading $\beta.$ As anticipated from the results in section \ref{ssec:Comparison Whole Domain} the nonlinear forecast outperforms the linear forecast also over the shrinking predictable region. Numerical values for these cases are given in Tables \ref{table:Corr Shrinking Domain} \& \ref{table:NMSE Shrinking Domain}. Owing to the discretization and attendant cut-off, the average correlation $\overline{\rho}$ at $t=0$ is not unity, nor is the average NMSE $\overline{\mathcal{E}}$ identically zero (cf.\ figure \ref{fig:FullRes-Rho_NMSE}). This is partly the effect of the ca.\ 5\% of total energy which is neglected when truncating at $k_0$ and $k_N$. It can also be observed that the increasing directional spreading leads to better performance of the nonlinear forecast; this may be attributed to the effect of directionality on the nonlinear correction \eqref{eq: discrete nonlinear dispersion}\cite{Stuhlmeier2019}.

\begin{figure*}
    \centering
    \includegraphics[width=1.0\linewidth]{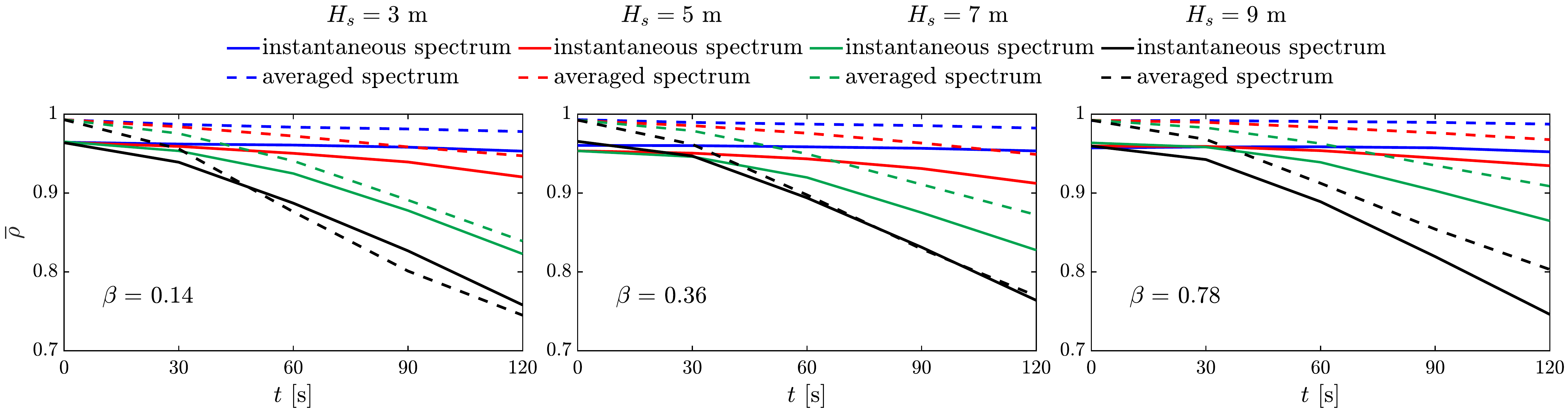}
    \caption{Comparison of average correlation $\bar{\rho}$ from nonlinear forecasts performed using a single-time amplitude spectrum (solid lines, ``instantaneous spectrum”) vs.\ forecasts using an averaged spectrum (dashed lines, ``averaged spectrum”) to calculate cut-offs $k_0, \, k_N$ and nonlinear frequency corrections $\Omega_r$.}
    \label{fig:rho_comparison_nonlinear_fc}
\end{figure*}

We note that, when the predictable region and dispersion correction \eqref{eq: discrete nonlinear dispersion} are computed using the ‘grassy’ Fourier amplitudes at $t=0$ s directly (cf.\ black and grey curves in Figure \ref{fig:Spectral-cut-offs-instantaneous-vs-averaged}), rather than calculating these from the power spectrum (blue, dashed curve in Figure \ref{fig:Spectral-cut-offs-instantaneous-vs-averaged}), the average correlation $\overline{\rho}$ between nonlinear forecast and HOS sea is slightly lower, as shown in Figure \ref{fig:rho_comparison_nonlinear_fc}. This figure compares the averaged correlation $\bar{\rho}$ between sea and nonlinear forecast when the nonlinear forecast is computed using either ``instantaneous spectrum" or ``averaged spectrum," as described in Section \ref{sec:Implementation}.  This discrepancy is due to the difficulty of defining stable cut-off values $k_0$ and $k_N$ with fluctuating instantaneous spectra. To a lesser degree it is attributable to the stability of the complex amplitudes \eqref{eq:B_p-Y-relation} which enter into the nonlinear frequency correction \eqref{eq: discrete nonlinear dispersion} when these are obtained from an averaged spectrum.

\section{Discussion \& Conclusions}
\label{sec:Discussion and Conclusions}

While basic ideas of deterministic forecasting have a long history \cite{sand1979three}, the computational effort associated with the production of forecasts in near real-time has constituted a major barrier to their practical application. Over the last two decades there has been a great deal of effort, particularly dedicated to accounting for nonlinear effects typical of real sea states, beginning with works of Morris et al. \cite{morris1998short} and Zhang et al. \cite{Zhang1999a}, the latter introducing a model that included nonlinear effects up to second order in wave steepness.

With a view to practical applications, our approach has been to extract the most critical feature of nonlinearity -- the dispersion correction -- from the initial conditions, and so to create a simple but robust improved forecast. We employ the FFT to extract modal amplitudes and phases at $t=0$ s. This initial data would furnish the material for a linear forecast; it can also be used to calculate the modal frequencies according to the third-order amplitude-dependent dispersion relation. In contrast with previous work \cite{Galvagno2021,Stuhlmeier2021}, we show that the amplitude dependence need not be computed from the Fourier modes known at time $t=0$ s, but should instead be extracted from the (averaged) energy spectrum. This is generally much more stable than the amplitude spectrum, and leads to a more robust prediction: it allows for a clearer identification of the energetically important components of the sea which determine the predictable region, and likewise avoids the need to recalculate the corrected frequencies for every prediction. These can instead be stored and updated only as needed.

In order to test the new forecasting methodology we have used the 5th order HOS \cite{Ducrozet2016} to generate synthetic sea-surfaces, ranging from short crested to long crested, and from low $(H_s/\lambda_p=1.9\%)$ to high $(H_s/\lambda_p=5.7\%)$ steepness. Both linear and nonlinear dispersion have been tested (on 10 realisations of each sea state), and the fidelity of the nonlinear forecast is shown to be consistently superior when measured using normalised mean square error as well as linear correlation. 
As the time-scale of nonlinear evolution is slow (of order $T/\epsilon^2$, for $T$ a typical period and $\epsilon$ a typical wave steepness), the difference in the forecasts is more pronounced at longer times ($t=90, \, 120$ s) and for steeper waves ($H_s = 7, \, 9$ m). Interestingly, we also find that the quality of the nonlinear forecast improves with increasing directional spreading, holding other values constant. It should be emphasised that all aspects of the nonlinear forecast are precomputed based on data available at $t=0$ s. Consequently the computational effort involved in the linear and nonlinear forecast is identical. For this simple reason we recommend the use of nonlinear dispersion corrections whenever phase-resolved predictions are needed.

In our treatment of the deterministic wave forecasting problem the starting point has been a perfectly accurate, synthetic snapshot of the sea surface. In practical applications such snapshots must be obtained instrumentally, and will contain errors as well as gaps whose characteristics depend on the instrumentation\cite{Fucile2016,Fucile2018}. For example, X-band radar \cite{reichert1999wamos} exhibits inevitable wave shadowing effects \cite{Belmont2023a,Belmont2023b}. Moreover antenna rotation means that measurements are typically not instantaneous, introducing additional complications \cite{Al-Ani2019}. While other technologies, such as stereo-imaging \cite{Bergamasco2017} or polarimetric imaging \cite{Ginio2023} may be able to provide virtually error-free reconstructions of the surface elevation, their range is typically limited and 
surface elevations are currently not available in real time using such techniques. One possibility is to employ data assimilation to reconstruct the free surface prior to making a prediction, as suggested by Desmars et al\cite{Desmars2022}, Qi et al\cite{Qi2018} or Naaijen et al\cite{Naaijen2018} who use sequences of images. Wang \& Pan \cite{Wang2021} have presented a framework which integrates data assimilation using ensemble Kalman filtering and HOS simulations for deterministic forecasting. Our work also assumes that no currents are present, which is an idealisation. Surface current data is obtainable from radar data \cite{STEWART19741039,Smeltzer2021,Lund2015,Smelzter2019}, and approximating this with a depth-averaged current can allow for this to be incorporated as a Doppler shift\cite{Fenton1985} in the dispersion relation. A holistic approach to data assimilation and forecasting, including simultaneous estimation of a (depth uniform but possibly unsteady) current, has recently been proposed by Wang et al\cite{Wang2022}.

Fourier transform techniques are by far the most commonly used method in processing the sea-surface elevation data. However, these too may introduce errors, particularly relating to end effects, as discussed by Hlophe et al\cite{Hlophe2021,Hlophe2022}. Other possibilities have also been explored, such as convolution of the measured waves with the sea's finite impulse response filter \cite{belmont2006filters}. When treating mixed space-time data, approaches which avoid Fourier transform methods have been proposed and tested by Al-Ani et al\cite{Al-Ani2019,Al-Ani2020}. While it may be possible to adapt our nonlinear dispersion correction to work with such methods, we have presented it here in terms of Fourier transforms for ease of applicability. 
In contrast to some current streams of deterministic forecasting research which favour numerical methods, e.g.\ employing the HOS to propagate a prediction forward in time, we have presented a lightweight approach easily substituted in place of the linear dispersion relation. We believe this can be profitably incorporated into other forecasting frameworks, where possible together with data assimilation, and ultimately feed into more accurate predictions of ship or device motions in steep sea-states. 
\section*{Data Availability Statement}
Data sharing is not applicable to this article as no new data were created or analyzed in this study.

\section*{Acknowledgements}
The authors would like to thank the anonymous reviewers of this manuscript for numerous helpful suggestions. RS and DA were funded by EPSRC Grant EP/V012770/1. RS gratefully acknowledges funding from an IMA QJMAM Fund Grant supporting research visits by MG to the University of Plymouth, and a UUKi UK-Israel Mobility Award supporting collaboration with Technion's Faculty of Civil \& Environmental Engineering. EM gratefully acknowledges the financial support by the scholarships from the Nancy and Stephen Grand Technion Energy Program and the funding by SolarEdge Technologies Ltd., as part of the Guy Sela Memorial Project at the Technion.
\clearpage

\appendix
\section{Tables: VALUES OF COMPARISON METRICS}
\label{sec:tables}
\subsection{Whole-domain forecasts}
The tables in this section show averaged correlations (Table \ref{table:Corr Full Domain}) and NMSE (Table \ref{table:NMSE Full Domain}) for linear/nonlinear forecasts vs HOS with identical resolution, at various values of $H_s$ and directional spreading angle $\beta$.
\begin{table}[h!]
\centering
\begin{tabular}{lllllll}
\toprule
	\multicolumn{7}{c}{$H_s = 3$ m} \\		
	\midrule			
 &	\multicolumn{2}{c}{$\beta=0.14$}	&	\multicolumn{2}{c}{$\beta=0.36$	} &	\multicolumn{2}{c}{$\beta=0.78$}	\\
$t$ [s]  & Lin &	Nonlin&	Lin&	Nonlin &	Lin&	Nonlin \\
30 &	0.9984&	0.9985&	0.9985	&0.9986	&0.9987&	0.9988 \\
60 &	0.9942&	0.9955&	0.9947&	0.9958	&0.995&	0.9966 \\
90	&0.9877	&0.9914	&0.9886	&0.9921	&0.9893	&0.9937 \\
120 &	0.9786&	0.9862&	0.9805&	0.9873	&0.9816&	0.9901 \\
\\
	\multicolumn{7}{c}{$H_s=5$ m}\\		
	\midrule			
 &	\multicolumn{2}{c}{$\beta=0.14$}	&	\multicolumn{2}{c}{$\beta=0.36$	} &	\multicolumn{2}{c}{$\beta=0.78$}	\\	
$t$ [s] & 	Lin &	Nonlin&	Lin&	Nonlin &	Lin&	Nonlin \\
$30$	&0.9920&	0.9917&	0.9110	&0.9918	&0.9920	&0.9930\\
$60$	&0.9660&	0.9721&	0.9637	&0.9725	&0.9667	&0.9777\\
$90$	&0.9245&	0.9444&	0.9199	&0.9460	&0.9272	&0.9578\\
$120$ &	0.8683&	0.9100&	0.8626&	0.9146&	0.8755&	0.9347\\
\\	
\bottomrule
\end{tabular} 
\end{table}
\newpage
\begin{table}
\centering
\begin{tabular}{lllllll}
\toprule
 \multicolumn{7}{c}{$H_s=7$ m}	\\
	\midrule				
 &	\multicolumn{2}{c}{$\beta=0.14$}	&	\multicolumn{2}{c}{$\beta=0.36$	} &	\multicolumn{2}{c}{$\beta=0.78$}	\\
$t$ [s] &	 Lin &	Nonlin&	Lin&	Nonlin &	Lin&	Nonlin \\
$30$	&0.9698&	0.9711&	0.9713	&0.9721	&0.9734	&0.9765\\
$60$	&0.8779&	0.9057&	0.8798	&0.9108	&0.8892	&0.9266\\
$90$	&0.7430&	0.8302&	0.7490	&0.8426	&0.7657	&0.8708\\
$120$ &	0.5862&	0.7639&	0.5989&	0.7857&	0.6157&	0.8233\\
\\
	\multicolumn{7}{c}{$H_s=9$ m}\\			
	\midrule			
 &	\multicolumn{2}{c}{$\beta=0.14$}	&	\multicolumn{2}{c}{$\beta=0.36$	} &	\multicolumn{2}{c}{$\beta=0.78$}	\\
$t$ [s] &	Lin &	Nonlin&	Lin&	Nonlin &	Lin&	Nonlin \\
$30$	&0.9252&	0.9260&	0.9295	&0.9301	&0.9321	&0.9395\\
$60$	&0.7142&	0.7844&	0.7298	&0.8089	&0.7408	&0.8360\\
$90$	&0.4624&	0.6756&	0.4821	&0.7248	&0.4959	&0.7625\\
$120$ &	0.2319&	0.6108&	0.2401&	0.6800	&0.2539	&0.7277\\
\bottomrule
\end{tabular}
\caption{Correlations between linear/nonlinear forecasts and HOS seas at various values of $H_s$ and directional spreading angle $\beta.$}
\label{table:Corr Full Domain}
\end{table}

\begin{table}
\centering
\begin{tabular}{lllllll}
\toprule
	\multicolumn{7}{c}{$H_s=3$ m} \\		
	\midrule			
 &	\multicolumn{2}{c}{$\beta=0.14$}	&	\multicolumn{2}{c}{$\beta=0.36$	} &	\multicolumn{2}{c}{$\beta=0.78$}	\\
$t$ [s]&	Lin &	Nonlin&	Lin&	Nonlin &	Lin&	Nonlin \\
30	& 0.0032& 0.0029& 0.0029& 0.0028& 0.0027& 0.0024 \\ 
60  & 0.0116& 0.0089& 0.0107& 0.0083& 0.0100& 0.0068 \\ 
90  & 0.0247& 0.0172& 0.0228& 0.0158& 0.0213& 0.0126 \\ 
120 & 0.0428& 0.0276& 0.0390& 0.0255& 0.0367& 0.0198 \\
						\\
	\multicolumn{7}{c}{$H_s=5$ m}\\		
	\midrule			
 &	\multicolumn{2}{c}{$\beta=0.14$}	&	\multicolumn{2}{c}{$\beta=0.36$	} &	\multicolumn{2}{c}{$\beta=0.78$}	\\	
$t$ [s]&	Lin &	Nonlin&	Lin&	Nonlin &	Lin&	Nonlin \\
30	& 0.0159& 0.0166& 0.0178& 0.0165& 0.0160& 0.0141 \\
60	& 0.0679& 0.0557& 0.0726& 0.0550& 0.0666& 0.0446 \\ 
90	& 0.1511& 0.1112& 0.1601& 0.1080& 0.1457& 0.0845 \\
120	& 0.2635& 0.1800& 0.2748& 0.1708& 0.2491& 0.1307 \\
\\	
\bottomrule
\end{tabular} 
\end{table}
\begin{table}[h!]
\centering
\begin{tabular}{lllllll}
\toprule
	\multicolumn{7}{c}{$H_s=7$ m}	\\
	\midrule				
 &	\multicolumn{2}{c}{$\beta=0.14$}	&	\multicolumn{2}{c}{$\beta=0.36$	} &	\multicolumn{2}{c}{$\beta=0.78$}	\\
$t$ [s]&	Lin &	Nonlin&	Lin&	Nonlin &	Lin&	Nonlin \\
30	& 0.0604& 0.0579& 0.0573& 0.0558& 0.0532& 0.0471 \\
60	& 0.2442& 0.1886& 0.2405& 0.1783& 0.2215& 0.1468 \\ 
90	& 0.5140& 0.3396& 0.5020& 0.3147& 0.4685& 0.2585 \\
120	& 0.8277& 0.4723& 0.8023& 0.4285& 0.7685& 0.3534 \\
\\
	\multicolumn{7}{c}{$H_s=9$ m}\\			
	\midrule			
 &	\multicolumn{2}{c}{$\beta=0.14$}	&	\multicolumn{2}{c}{$\beta=0.36$	} &	\multicolumn{2}{c}{$\beta=0.78$}	\\
$t$ [s]&	Lin &	Nonlin&	Lin&	Nonlin &	Lin&	Nonlin \\
30	& 0.1496& 0.1481& 0.1410& 0.1397& 0.1358& 0.1210 \\
60	& 0.5717& 0.4311& 0.5404& 0.3823& 0.5183& 0.3280 \\
90	& 1.0755& 0.6489& 1.0358& 0.5503& 1.0080& 0.4750 \\
120	& 1.5365& 0.7785& 1.5197& 0.6400& 1.4920& 0.5446 \\
\bottomrule
\end{tabular}
\caption{NMSE for linear/nonlinear forecasts and HOS seas at various values of $H_s$ and directional spreading angle $\beta.$}
\label{table:NMSE Full Domain}
\end{table}

\clearpage
\subsection{Shrinking domain forecasts}
The tables in this section show averaged correlations (Table \ref{table:Corr Shrinking Domain}) and NMSE (Table \ref{table:NMSE Shrinking Domain}) for linear/nonlinear forecasts vs HOS within the predictable region, at various values of $H_s$ and directional spreading angle $\beta$.
\begin{table}[h!]
\centering
\begin{tabular}{lllllll}
\toprule
	\multicolumn{7}{c}{$H_s=3$ m} \\		
	\midrule			
 &	\multicolumn{2}{c}{$\beta=0.14$}	&	\multicolumn{2}{c}{$\beta=0.36$	} &	\multicolumn{2}{c}{$\beta=0.78$}	\\
$t$ [s]&	Lin &	Nonlin&	Lin&	Nonlin &	Lin&	Nonlin \\
$30$ & 0.9867& 0.9870& 0.9890& 0.9895& 0.9913& 0.9918 \\
$60$& 0.9813& 0.9835& 0.9846& 0.9874& 0.9879& 0.9907 \\
$90$  & 0.9746& 0.9812& 0.9784& 0.9855& 0.9826& 0.9896 \\
$120$ & 0.9647& 0.9780& 0.9692& 0.9824& 0.9746& 0.9874 \\
						\\
	\multicolumn{7}{c}{$H_s=5$ m}\\		
	\midrule			
 &	\multicolumn{2}{c}{$\beta=0.14$}	&	\multicolumn{2}{c}{$\beta=0.36$	} &	\multicolumn{2}{c}{$\beta=0.78$}	\\	
$t$ [s]&	Lin &	Nonlin&	Lin&	Nonlin &	Lin&	Nonlin \\
$30$	& 0.9815& 0.9840& 0.9813& 0.9854& 0.9855& 0.9897 \\
$60$	& 0.9577& 0.9723& 0.9564& 0.9759& 0.9627& 0.9833 \\
$90$	& 0.9205& 0.9585& 0.9161& 0.9634& 0.9250& 0.9763 \\
$120$ & 0.8815& 0.9472& 0.8640& 0.9490& 0.8727& 0.9677 \\
\\				
\bottomrule
\end{tabular} 
\end{table}
\begin{table}
\centering
\begin{tabular}{lllllll}
\toprule
	\multicolumn{7}{c}{$H_s=7$ m}	\\
	\midrule				
 &	\multicolumn{2}{c}{$\beta=0.14$}	&	\multicolumn{2}{c}{$\beta=0.36$	} &	\multicolumn{2}{c}{$\beta=0.78$}	\\
$t$ [s] &	Lin &	Nonlin&	Lin&	Nonlin &	Lin&	Nonlin \\
$30$	& 0.9640& 0.9755& 0.9647& 0.9790& 0.9682& 0.9829 \\
$60$	& 0.8822& 0.9407& 0.8787& 0.9497& 0.8923& 0.9628 \\
$90$	& 0.7561& 0.8910& 0.7457& 0.9109& 0.7668& 0.9349 \\
$120$ & 0.6145& 0.8390& 0.5933& 0.8724& 0.6364& 0.9086 \\
\\
	\multicolumn{7}{c}{$H_s=9$ m}\\			
	\midrule			
 &	\multicolumn{2}{c}{$\beta=0.14$}	&	\multicolumn{2}{c}{$\beta=0.36$	} &	\multicolumn{2}{c}{$\beta=0.78$}	\\
$t$ [s]&	Lin &	Nonlin&	Lin&	Nonlin &	Lin&	Nonlin \\
$30$	& 0.9239& 0.9560& 0.9301& 0.9617& 0.9299& 0.9680 \\
$60$	& 0.7502& 0.8762& 0.7521& 0.8977& 0.7400& 0.9124 \\
$90$	& 0.5535& 0.8011& 0.5279& 0.8290& 0.5057& 0.8542 \\
$120$ & 0.3485& 0.7451& 0.3112& 0.7695& 0.2710& 0.8031 \\
\bottomrule
\end{tabular}
\caption{Correlations between linear/nonlinear forecasts and HOS seas at various values of $H_s$ and directional spreading angle $\beta$, for predictions of shrinking regions.}
\label{table:Corr Shrinking Domain}
\end{table}
\newpage
\begin{table}
\centering
\begin{tabular}{lllllll}
\toprule
	\multicolumn{7}{c}{$H_s=3$ m} \\		
	\midrule			
 &	\multicolumn{2}{c}{$\beta=0.14$}	&	\multicolumn{2}{c}{$\beta=0.36$	} &	\multicolumn{2}{c}{$\beta=0.78$}	\\
$t$ [s]&	Lin &	Nonlin&	Lin&	Nonlin &	Lin&	Nonlin \\
$30$	& 0.0266& 0.0260& 0.0222& 0.0210& 0.0174& 0.0163 \\ 
$60$	& 0.0376& 0.0331& 0.0310& 0.0254& 0.0242& 0.0185 \\ 
$90$	& 0.0515& 0.0381& 0.0434& 0.0290& 0.0345& 0.0207 \\ 
$120$	& 0.0718& 0.0450& 0.0617& 0.0354& 0.0506& 0.0252 \\
						\\
	\multicolumn{7}{c}{$H_s=5$ m}\\		
	\midrule			
 &	\multicolumn{2}{c}{$\beta=0.14$}	&	\multicolumn{2}{c}{$\beta=0.36$	} &	\multicolumn{2}{c}{$\beta=0.78$}	\\	
$t$ [s]&	Lin &	Nonlin&	Lin&	Nonlin &	Lin&	Nonlin \\
$30$	& 0.0371& 0.0320& 0.0376& 0.0292& 0.0288& 0.0205 \\
$60$	& 0.0851& 0.0557& 0.0875& 0.0484& 0.0742& 0.0333 \\ 
$90$	& 0.1621& 0.0846& 0.1683& 0.0735& 0.1493& 0.0472 \\
$120$	& 0.2419& 0.1082& 0.2745& 0.1026& 0.2541& 0.0645 \\
\\	
\bottomrule
\end{tabular} 
\end{table}
\begin{table}
\centering
\begin{tabular}{lllllll}
	\multicolumn{7}{c}{$H_s=7$ m}	\\
	\midrule				
 &	\multicolumn{2}{c}{$\beta=0.14$}	&	\multicolumn{2}{c}{$\beta=0.36$	} &	\multicolumn{2}{c}{$\beta=0.78$}	\\
$t$ [s]&	Lin &	Nonlin&	Lin&	Nonlin &	Lin&	Nonlin \\
$30$	& 0.0719& 0.0490& 0.0707& 0.0422& 0.0633& 0.0341 \\
$60$	& 0.2362& 0.1189& 0.2443& 0.1011& 0.2144& 0.0741 \\ 
$90$	& 0.4905& 0.2200& 0.5120& 0.1789& 0.4653& 0.1297 \\
$120$	& 0.7780& 0.3260& 0.8154& 0.2560& 0.7172& 0.1812 \\
\\
	\multicolumn{7}{c}{$H_s=9$ m}\\			
	\midrule			
 &	\multicolumn{2}{c}{$\beta=0.14$}	&	\multicolumn{2}{c}{$\beta=0.36$	} &	\multicolumn{2}{c}{$\beta=0.78$}	\\
$t$ [s]&	Lin &	Nonlin&	Lin&	Nonlin &	Lin&	Nonlin \\
$30$	& 0.1526& 0.0881& 0.1402& 0.0768& 0.1393& 0.0636 \\
$60$	& 0.4990& 0.2471& 0.4963& 0.2051& 0.5171& 0.1745 \\
$90$	& 0.8973& 0.3982& 0.9429& 0.3417& 0.9834& 0.2899 \\
$120$	& 1.3272& 0.5183& 1.3772& 0.4602& 1.4082& 0.3824 \\
\bottomrule
\end{tabular}
\caption{NMSE for linear/nonlinear forecasts and HOS seas at various values of $H_s$ and directional spreading angle $\beta$, for predictions of shrinking regions.}
\label{table:NMSE Shrinking Domain}
\end{table}
\newpage
\clearpage
\bibliography{library,citations}
\end{document}